\documentclass[aps,twocolumn,showpacs,fleqn]{revtex4}

\usepackage{amsmath,amssymb}
\usepackage{graphicx}

\renewcommand{\vec}{\mathbf}

\def\Fig#1{Figure~\ref{#1}}
\def\fig#1{Fig.\,\ref{#1}}

\def\eq#1{Eq.\,(\ref{#1})}

\def\be#1{\begin{equation}\label{#1}}
\def\ee{\end{equation}}
\def\nmax{n_{\rm max}}

\DeclareMathOperator*{\sumint}{%
\mathchoice%
  {\ooalign{$\displaystyle\sum$\cr\hidewidth$\displaystyle\int$\hidewidth\cr}}
  {\ooalign{\raisebox{.14\height}{\scalebox{.7}{$\textstyle\sum$}}\cr\hidewidth$\textstyle\int$\hidewidth\cr}}
  {\ooalign{\raisebox{.2\height}{\scalebox{.6}{$\scriptstyle\sum$}}\cr$\scriptstyle\int$\cr}}
  {\ooalign{\raisebox{.2\height}{\scalebox{.6}{$\scriptstyle\sum$}}\cr$\scriptstyle\int$\cr}}
}

\def\fig#1{Fig.\,\ref{#1}}
\def\eps{\varepsilon}
\def\e#1{{\rm e}^{#1}}
\def\Int#1#2#3{\int_{#1}^{#2}\!\!\!{\rm d}#3\,}
\def\jnt#1{\int\!{\rm d}#1\,}

\def\g{{\rm b}}
\def\i{{\rm i}}
\def\e#1{{\rm e}^{#1}}

\begin{document}

\title{The envelope Hamiltonian for electron interaction with ultrashort pulses}
\author{Koudai Toyota}\author{Ulf Saalmann}\author{Jan M. Rost}
\affiliation{Max Planck Institute for the Physics of Complex Systems\\
 N\"othnitzer Stra{\ss}e 38, 01187 Dresden, Germany }

\begin{abstract}\noindent
For ultrashort VUV pulses with a pulse length comparable to the orbital time of the bound electrons they couple to we propose a simplified envelope Hamiltonian. It is based on the Kramers-Henneberger representation in connection with a Floquet expansion of the strong-field dynamics but keeps the time dependence of the pulse envelope explicit. Thereby, the envelope Hamiltonian captures the essence of the physics\,---\,light-induced shifts of bound states, single-photon absorption, and non-adiabatic electronic transitions. It delivers quantitatively accurate ionization dynamics and allows for physical insight into the processes occurring. Its minimal requirements for construction in terms of laser parameters make it ideally suited for a large class of atomic and molecular problems.
\end{abstract}

\maketitle

\noindent
Interaction of strong light fields with bound electrons continues to produce new experimental phenomena while theory is looking for appropriate approximations since this dynamics, even for a single active electron, is not analytically solvable. The interaction with femtosecond pulses in the infrared domain is by now well understood. Nonlinear mechanisms lead to high-harmonic generation (HHG) and above-threshold ionization (ATI) with photon and electron emission at high energies \cite{leba+94,pabe+94}. In contrast, the domain of slow photo-electrons has only become recently a center of attention through the discovery of substantial photo-electron yields with a couple of eV kinetic energy in atoms and molecules exposed to mid-infrared laser fields \cite{blca+09,quli+09}. These findings have triggered intensive research to identify the mechanism behind this phenomenon which was finally found in soft recollisions \cite{quli+09,kasa+12,kasa+12a,lebu+13} complementing the hard recollisions inducing HHG and ATI.

In parallel advanced optical techniques allow for the generation of attosecond light pulses so short that their pulse duration $T$ can reach the period of a bound electron $T_\nu$ they couple to. Also here, in simulations a surprisingly large yield of low energy electrons was found \cite{toto+09} despite the fact that the carrier frequency $\omega$ was high enough to elevate the photo electrons well into the continuum by single-photon ionization (SPI). The latter is a realistic scenario for such pulses, since the carrier frequency is high enough to fit a few cycles into the pulse and therefore able to provide a well-defined main energy of the photons. At the same time such pulses, although quite strong in terms of absolute intensity do not contain too many photons due to the short pulse length and remain well in the non-relativistic domain of electron dynamics. 
Acknowledging the fact that for this regime of light-matter coupling, characterized by the hierarchy of time scales 
\be{time_hierarchy}
T \sim T_\nu > T_\omega\,
\ee
with $T_\omega = 2\pi/\omega$, the time envelope of the laser pulse (characterized by $T$) becomes dynamically important.
In the following we will formulate the envelope Hamiltonian which captures surprisingly accurately the ionization dynamics as will be demonstrated in comparison with the numerical solution of the time-dependent Schr\"odinger equation (TDSE) for two examples. We use atomic units unless stated otherwise and start from the Hamiltonian in the Kramers-Henneberger (KH) frame \cite{he68} \be{ham}
H = -\frac 12\boldsymbol{\nabla}^2 + V\big(\vec r \,{+}\, \vec{e}_{x}\,x_\omega(t)\big),
\ee
where
$V$ is the potential in which the electron is bound and $x_\omega$
is the classical quiver position in a linearly polarized laser field along $x$, marked by the unit vector $\vec{e}_{x}$.

To facilitate the analytical derivations leading to the envelope Hamiltonian, we 
define the laser field $F(t)$ in terms of the quiver amplitude entering \eq{ham},
\be{electric}
F(t) = - \frac{d^2x_\omega}{dt^2}
\ee
with $x_{\omega}$ specified analytically as
\begin{subequations}\label{quiver}\begin{align}
x_\omega(t) & = \alpha(t)\cos(\omega t +\delta),\label{quiver-a}
\\
\alpha(t) & \equiv\alpha_0 \,\e{-4\ln2(t/T)^2}.\label{quiver-b}
\end{align}\end{subequations}
Thus $F(t)$ describes a finite pulse with duration $T$ (full width at half maximum), and it represents a proper electromagnetic wave with vanishing DC component $\int\!{\rm d}t\, F(t) = dx_\omega(t)/dt|^{+\infty}_{-\infty}=0$.
For the pulse to remain in the non-relativistic domain, we characterize it with the maximum field strength $F(0) = F_{0}\cos\delta$ which leads to the prefactor 
\be{alfa0}
\alpha_0 = \frac{F_0}{\omega^2}\frac{1}{1+8\ln2/(T\omega)^2}
\ee 
in \eq{quiver-b} following directly from \eq{electric}.

While the Hamiltonian \eq{ham} contains of course all dynamics implicitly, we aim at a formulation
which brings out explicitly the relevant physical processes\,---\,light-induced shifts of the bound states, and non-adiabatic as well as $n$-photon induced electron transitions. We first construct a Hamiltonian, which is formally exact in the limit $\nmax\to \infty$,
\begin{subequations}\be{hred}
H_{\nmax}(t) = -\frac{1}{2}\boldsymbol\nabla^{2} +\!\!\!\!\!\sum_{n=-\nmax}^{+\nmax}\!\!\!\! V_{n}(\vec r,t)\,\e{-\i n\omega t},
\ee
where the $V_n(\vec r, t)$ 
are \emph{single-cycle} averaged Fourier-components of the potential in \eq{ham},
\be{Fcomponent}
V_n(\vec r,t) = \frac{1}{T_\omega}\Int{0}{T_{\omega}}{t'} V\big(\vec r+\vec{e}_{x}\,\alpha(t)\cos(\omega t'+\delta)\big)\e{\i n\omega t'}\,.
\ee\end{subequations}
 Note, that \eq{hred} is {\it not} the usual Floquet representation which would absorb the entire time dependence in the Fourier phases $e^{\i n\omega t}$ with time-independent coefficients $V_n = V_{n}(\vec r)$. We define $H_{2}$ with an expansion length
 of $\nmax=2$ (maximally two-photon exchange) as the envelope Hamiltonian, since it agrees per constructionem for small $\alpha_0$ with the KH Hamiltonian of \eq{ham}, see supplemental material \cite{supple}. This means, that the envelope Hamiltonian $H_{2}$ is exact
in the two extreme and seemingly opposite limits, namely for short pulse length $T$ and for large photon frequency/very short optical period $T_{\omega}$ which implies $T\gg T_{\omega}$, see \eq{alfa0}.
A reflection on the building blocks of $H_{2}$ reveals that it should be also valid for finite pulses and frequencies, as long as
one photon takes a bound electron into the continuum. This is indeed the case and will be demonstrated by formulating 
 an adiabatic time-dependent perturbation theory (aTDPT) for time dependent basis functions to work out the mechanisms which lead to ionization with \eq{hred}.
In fact, despite the presence of a strong field in terms of the ponderomotive potential the $V_{n}, |n|\ge 1$ can be treated as perturbations for suitable initial and final states, and
aTDPT gives itself accurate quantitative results and simplifies the treatment of short pulses beyond the solution of $H_{2}$ since it requires only the solution of 
\be{ham0}
H_{0}(t) = -\frac 12 \boldsymbol{\nabla}^2 + V_{0}(\vec r, t)\,,
\ee
while the other terms $V_{n}$ in 
\eq{hred} can be treated perturbatively.
We expand the wave function 
into eigenstates $|\beta(t)\rangle$ of $H_{0}(t)$ at fixed $t$, 
\be{wf}
|\psi\rangle = e^{-\i\chi(t)}\sumint c_{\beta}(t)|\beta(t)\rangle\,e^{-\i t E_{\beta}(t)}\,,
\ee
where $|\beta(t)\rangle$ denotes both, bound and continuum eigenstates.
 The phase $\chi(t)$ reflects the fact that eigenstates are defined up to a phase which can be time-dependent in our case and which is used to simplify the coupled differential equations for the coefficients $c_{\beta}(t)$. Details can be found in the supplement \cite{supple}. 
In 1st-order time-dependent perturbation theory of the coefficients $c_{\beta}$
and with initially only the bound state $|\g(t)\rangle$ occupied ($c_{\g}^{(0)}(t)=1$, $c_{\beta\ne b}^{(0)}(t) = 0$) 
the differential photo-ionization probability is given as usual by $\frac{dP}{d\vec k} = \lim_{t\to\infty}|c^{(1)}_{\vec k}(t)|^{2}$. With the envelope Hamiltonian $H_{2}$
the contributions to the cross section
\begin{subequations}
\label{1storder}
\be{photoion}
\frac{dP}{d\vec k} = \Big|{\textstyle\sum\limits_{n=-2}^{2}}\lim_{t\to\infty}M_{n}(\vec k,t)\Big|^{2}
\ee
can be disentangled according to the number of photons $n$ exchanged with $H_{0}(t)$ and
described by the matrix elements 
\be{a0}
 M_{0}(\vec k,t) = -\Int{-\infty}{t}{t'}%
\Big\langle \vec k,t'\Big|\frac{\partial}{\partial t'}\Big|\g(t') \Big\rangle
\e{\i\phi_{0}(t')}
\ee
and for one- and two-photon transitions, $n\,{=}\,{\pm}1,\,{\pm}2$,
\be{an}
M_{n}(\vec k,t) = -\i\Int{-\infty}{t}{t'}
\Big\langle \vec k,t'\Big|V_{n}(x,t')\Big|\g(t') \Big\rangle\e{\i\phi_{n}(t')}
\ee
with the phases
\be{phase}
\phi_{n}(t) = [k^{2}/2- E_{\g}(t)-n\omega]t\,.
\ee
\end{subequations}
While \eq{1storder} looks familiar on a first glance, it exhibits clear differences compared to 
standard TDPT: Without an explicit time-dependence of the basis functions $|\beta(t)\rangle$, the matrix
element for non-adiabatic transitions \eq{a0} vanishes. A bit more subtle is the energy difference in the 
phases $\phi_{n}$: While the energy characterizing the final ionized state is not time-dependent, $E_{
\beta}=\eps_{k}=k^{2}/2$,
the energy of the initial bound state is actually a time average $ E_{\g}(t) = t^{-1}\int^{t}\!{\rm d}t'\,\eps_{\g}(t')$ while 
$\big[H_{0}(t) -\eps_\g(t)\big]|\rm b(t)\rangle = 0$.

\begin{figure}[t]
\centerline{\includegraphics[scale=0.55]{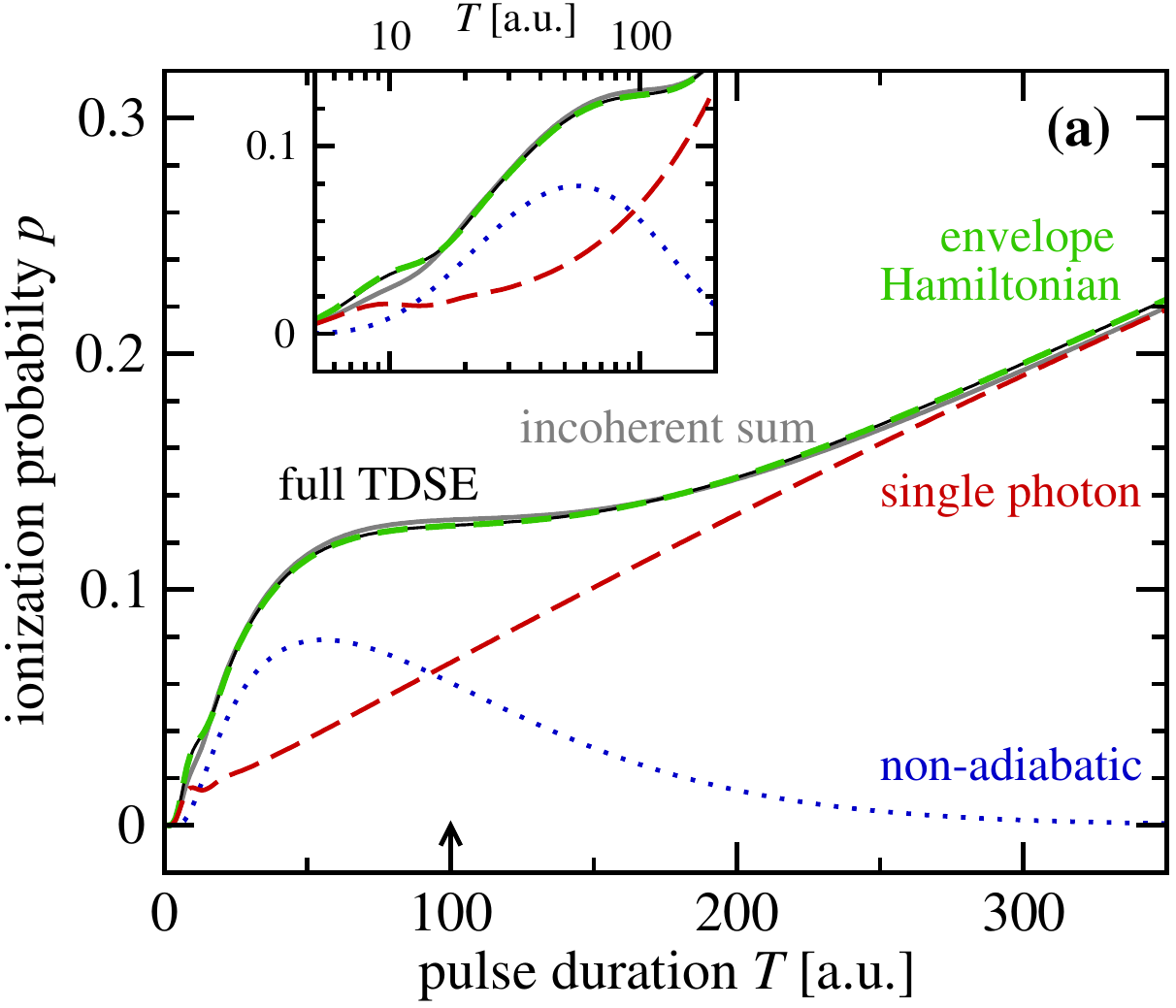}}
\centerline{\includegraphics[scale=0.55]{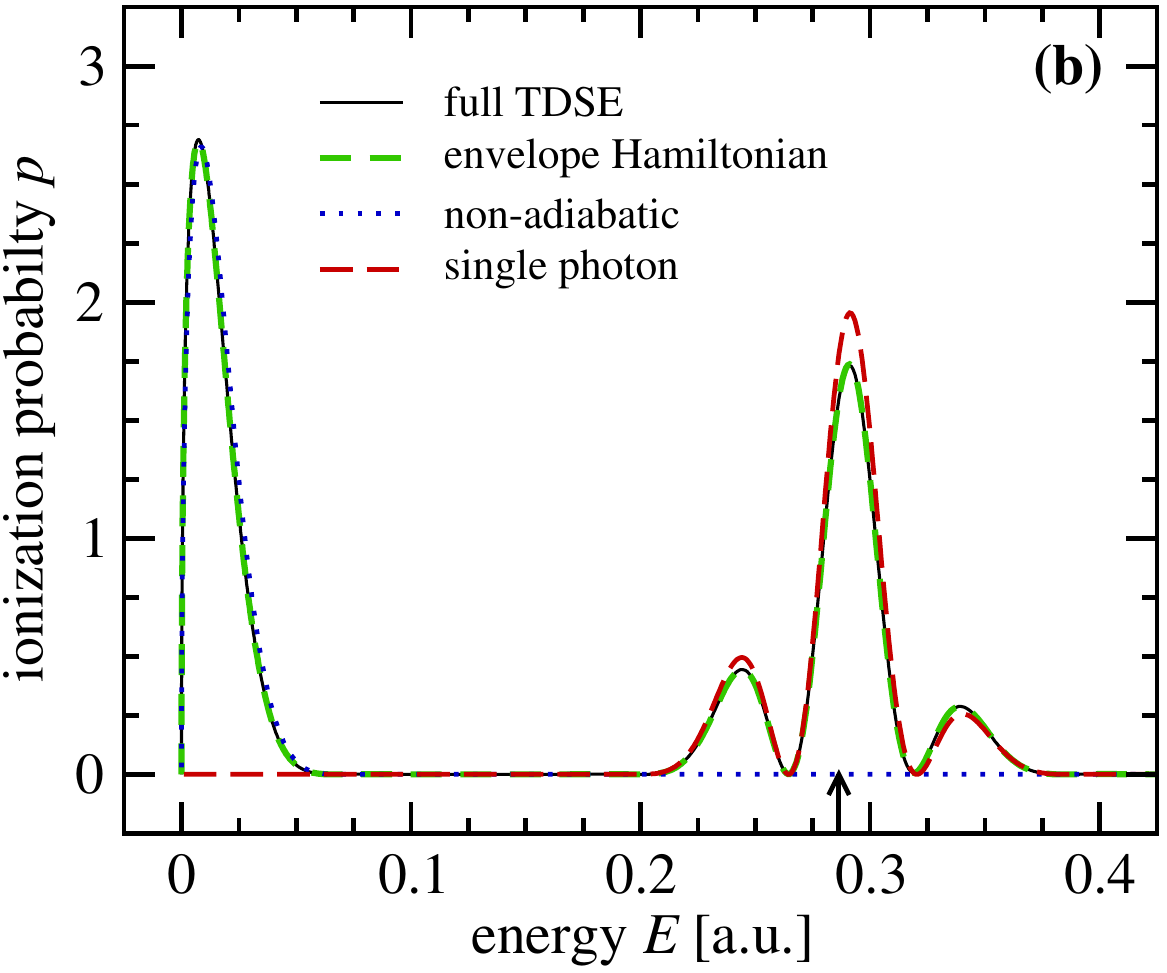}}
\caption{Ionization probability (a) $P(T) = \int\!{\rm d}E\,dP/dE$ as function of pulse length $T$ and (b) $dP/dE$ as a function of photo-electron energy at $T=100\,$au. Results are shown 
for the KH Hamiltonian \eq{ham} with $V(x)$ from \eq{potm} (black-solid) and for the envelope Hamiltonian $H_{2}$ according to \eq{hred} with the same potential (green-dashed), as well as for partial contributions from non-adiabatic transitions $P_{0}$ (blue-dotted) and from single-photon transitions $P_{1}$ (red-dashed) as well as their sum $P_{0}+P_{1}$ (grey-solid). The arrow in (a) points to the pulse duration shown in (b). The one in (b) points to the SPI energy $\eps_\g +\omega$ of the unperturbed system. The laser pulse parameters are $F_0 = 0.5$ and $\omega = 0.314$.
}
\label{fig:ionp}
\end{figure}

The mechanisms behind the different contributions $M_{n}$ to the cross section can be best explained with an example. To this end we will discuss a one-dimensional problem for a negative ion \cite{greb93} with the model potential 
\be{potm}
V(x) = - \frac{\exp[-a_{1}\sqrt{(x/a_{1})^2 + a_{2}{\!}^2}]}{
\sqrt{(x/a_{1})^2+a_{3}{\!}^2}}
\ee
with $a_{1} = 24.856$, $a_{2} = 0.16093$ and $a_{3} = 0.25225$. With these parameters
$V(x)$ supports one bound state of energy $E_\g = -0.0277\,$au.
 \Fig{fig:ionp}a shows the ionization probability as a function of pulse length with almost perfect agreement between the full dynamics (black-solid) and the one obtained with the envelope Hamiltonian $H_{2}$ (green-dashed).
 This is particularly astonishing for pulse lengths around the period of the bound electron, $T_{\g}=69.1$\,au \footnote{We determine $T_{\g}$ from the classical period at the quantized energy $\eps_{\g}$ of the respective bound state, $T_{\g}= \oint{\rm d}x\,(2m/[\eps_{\g}-V(x)])^{1/2}$.}
which represents exactly the dynamical regime we are targeting since the optical period is $T_\omega = 20\,$au, cf.\ \eq{time_hierarchy}. The plateau structure, although strongly deviating from the SPI (red-dashed), is faithfully described by the envelope Hamiltonian. 
The photo-electron spectrum in \fig{fig:ionp}b represents a more differential property\,---\,yet the same good agreement between the full dynamics and that of the envelope Hamiltonian can be seen. 

The non-adiabatic contribution to the ionization probability $P_0 \equiv \int \frac{{\rm d}k}{2\pi}\, |M_{0}(k,\infty)|^{2}$, is shown in \fig{fig:ionp} (blue-dotted line).
It becomes substantial for $T<T_\g$, i.e., when the pulse length is shorter than the period of the electron. As a consequence, the electron cannot follow the pulse anymore and undergoes non-adiabatic transitions leading to accumulation of amplitude in the continuum. On the other hand for $T>T_\g$ the electron dynamics is adiabatic and $M_{0}$ does not lead to ionization. The SPI described by $ p_1 =\int \frac{{\rm d}k}{2\pi}\, \big|M_{1}(k,\infty)\big|^2$ is due to $V_{1}$ and grows linearly for large $T$ in accordance with a single-photon process. The total ionization probability reached is no longer small compared to unity and depletion of the ground state must be taken into account, 
leading to $P_{1}(T)=1 - \exp(-p_{1}(T))$ shown as red-dashed line. 
Remarkably, the {\it incoherent} sum of both probabilities $P_0 + P_1$ (grey-solid line in \fig{fig:ionp}a) approximates very accurately the full ionization probability $P(T)$ from the coherent superposition of amplitudes \eq{photoion} (green-dashed in \fig{fig:ionp}a).
Typically, this is the case if the differential probabilities $dP_n/dk=|M_{n}(k,\infty)|^{2}$ peak at very different momenta $k$ such that the overlap integrals $\jnt{k} M^*_{n}(k,\infty)M_{m}(k,\infty)$ for $n\ne m$ vanish.

This is indeed the case, as the photo-electron spectrum in \fig{fig:ionp}b reveals: non-adiabatic transitions produce low-energy electrons \cite{baor+99} whose probability
 $dP_{0}/dk$ (blue-dotted) is well separated from the contribution $dP_{1}/dk$ (red-dashed) mainly due to single-photon absorption. The latter peaks close but not at the CW laser photo-line at $E = \eps_\g(-\infty)+\omega= 0.2864$. This is the result of two competing effects. Firstly, there is a red-shift
 due to the finite width of the SPI peak which is convoluted with the exponentially decreasing dipole matrix element in \eq{an}. Secondly, there are the well-known light-induced shifts of the energy levels, notably the initial bound state $\eps_{\g}$ gets lifted upwards leading to a blue-shift of the SPI peak. In the limit of $\omega T\gg 1$ the red-shift disappears, while the blue-shift is only influenced by the intensity of the laser pulse.
 
\begin{figure}[b]
\centerline{\includegraphics[width=\columnwidth]{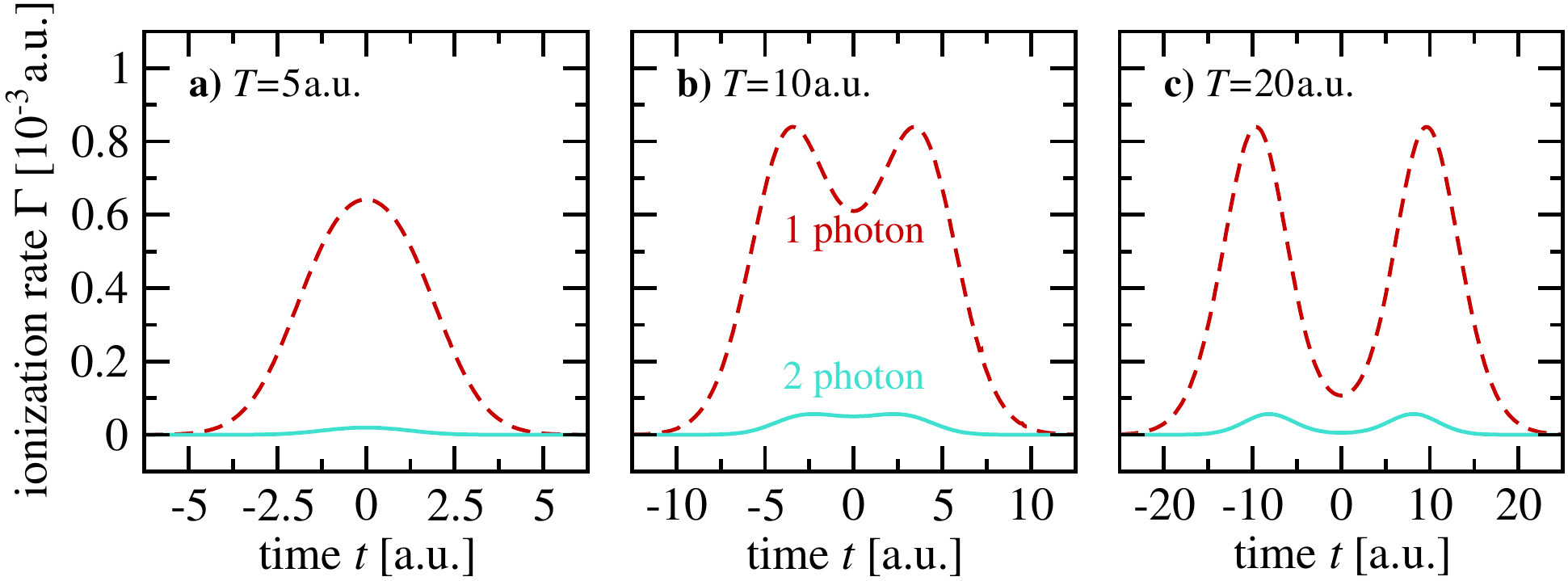}}
\caption{Photo-ionization probabilities per optical cycle $\Gamma_{n}(t)=T_{\omega}^{-1}\int \frac{dk}{2\pi}|\int_{0}^{T_{\omega}}M_{n}(k,t')dt'|^{2}$ as a function of time from the envelope Hamiltonian $H_{2}$ for a pulse length of $T=20$, for details see the supplement \cite{supple}.
}
\label{fig:V1V2}
\end{figure}%
However, the SPI peak does not only shift if generated by an intense short pulse, it also gets modulated as a result of interference of photo-electron emission in the rising and falling wing of the pulse \cite{toto+07,dece12} recently also found in molecules \cite{yufu+13}. The reason for this peculiar behavior can be interpreted as the onset of stabilization during the laser pulse revealed by the time dependence of the one- and two-photon transition probabilities $M_{n}$, $n =1,2$. For sufficiently large intensities (varied in \fig{fig:V1V2} with the help of the pulse length, see \eq{alfa0}) a double-hump structure appears corresponding to two distinct maximal ionization probabilities before and after the pulse maximum. The intermittent decreases of the ionization probability leading to this double hump is a signature of stabilization around the maximum of the laser pulse. It affects single- as well as multi-photon ionization as can be seen in \fig{fig:V1V2} but requires a large enough pulse amplitude and therefore does not occur for $T=5$. \Fig{fig:V1V2} also illustrates, why
the envelope Hamiltonian is such a good approximation: Two photon ionization (green line) does happen, but is already quite small, such that $n$-photon absorption with $n\,{\ge}\,3$ can be neglected. The reason is not a weak field\,---\,in fact the ponderomotive energy for $T$ being large is with $0.63$\,au quite large compared to the binding energy $E_{\g}\,{\approx}\,0.03$\,au. Small, however, is the probability for absorption of a (subsequent) photon in the continuum which is required for multi-photon absorption.
This explains, why terms $V_{n}$ up to $n=2$ are sufficient to capture the full dynamics. Of course, separating off the time dependence of the laser envelope is crucial: If this is not done, all time dependence is contained in the Fourier amplitudes $e^{\i n\omega t}$. In this case an expansion to $n$ much larger than the number of absorbed photons is necessary to capture a fast changing envelope \cite{mori+11}.

While the modulation described occurs due to interference within the single-photon contribution $|M_{1}|^{2}$, there 
is also an effect due to the coherence between the SPI and the non-adiabatic amplitude, but 
 only for pulse lengths comparable or shorter than the optical period $T_\omega$, where the SPI peak becomes very broad and starts to overlap with the non-adiabatic electron peak (see inset of \fig{fig:ionp}a).

Much more surprising is probably that the envelope Hamiltonian delivers such a good description of the dynamics for pulses with $T/T_{\omega}\,{\sim}\,1$ and $\alpha_{0}\,{>}\,1$, 
where neither the small $\alpha_{0}$ expansion can justify the envelope Hamiltonian, nor an adiabatic variation of the pulse envelope $(T/T_{\omega} \gg 1)$. For the parameters of \fig{fig:ionp}a with $F_{0}=0.5$ and $T_{\omega} = 20$ this situation corresponds to pulse lengths from about 3\,au ($\alpha_{0}{=}1.12$) to 200\,au ($T/T_{\omega}{\sim}10$).

\begin{figure}[t]
\centerline{\includegraphics[width=.8\columnwidth]{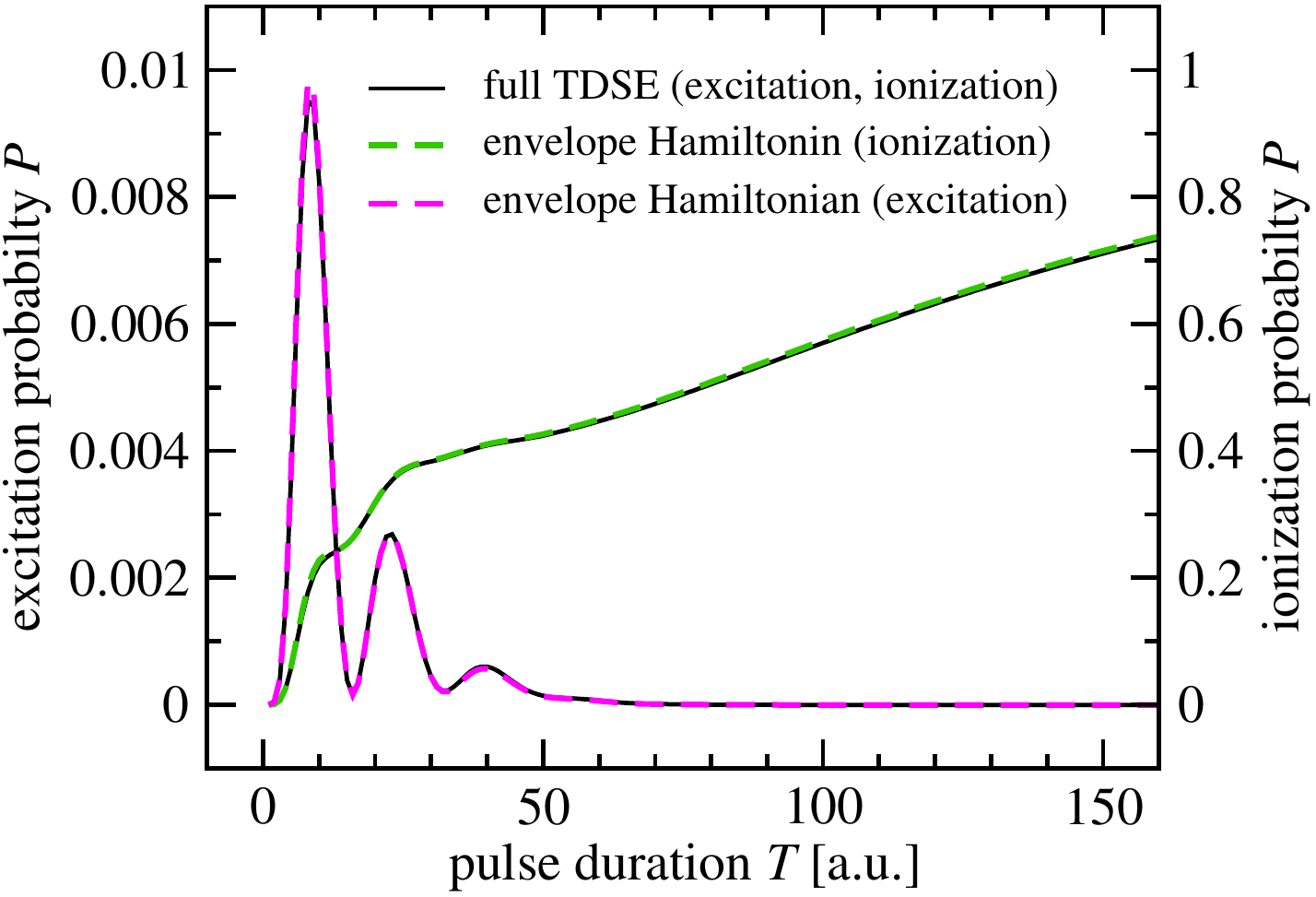}}
\caption{Photo-ionization and excitation using the potential $W(x) = -W_{0}\exp(-(x/\sigma)^{2})$
with $W_{0}=0.2$ and $\sigma = 2.65$ and the same pulse parameters as in \fig{fig:ionp}. Results obtained with the envelope Hamiltonian $H_{2}$ (dashed lines) and the full Hamiltonian (solid) are given as a function of pulse length $T$.
}
\label{fig:ionpgaussg}
\end{figure}%
One may of course ask, if the excellent quantitative and qualitative description of the short-pulse ionization dynamics provided by the envelope Hamiltonian is restricted to the potential \eq{potm} for which it has been demonstrated here, and more generally, to one-dimensional spatial dynamics. Clearly, the formulation of the envelope Hamiltonian does not make any use of dimensionally restricted dynamics. Moreover, we have seen that the incoherent superposition of non-adiabatic and SPI probabilities relies on the fact that both processes peak at different photo-electron momenta rendering the overlap integral of the respective amplitudes small. This effect will be eventually amplified but certainly not diminished when full 3D dynamics is considered. Less obvious is how the agreement depends on the form of the potential.
Therefore, we present below the result for a Gaussian binding potential of the form $W(x) = -W_{0}\exp(-(x/\sigma)^{2})$ whose parameters with a ground state of $E_\g= -0.1$ and corresponding electron period of $T_\g=34.1$
have been chosen such that the hierarchy of time scales 
\eq{time_hierarchy} is fulfilled with the same laser pulse we have been using for $V(x)$ from \eq{potm}. The potential $W$ supports in contrast to $V$ also an excited state. \Fig{fig:ionpgaussg} shows the probability for ionization and excitation. Again, the description with the envelope Hamiltonian is almost perfect. We have done similar calculation for other scenarios with short range Hamiltonians always finding excellent agreement with the full numerical solution. We expect our approach to be equally accurate for long range potentials.

So far the time-dependence for the envelope of the laser pulse in a Fourier representation of the KH Hamiltonian has been proposed only for adiabatically slowly varying envelopes in the literature \cite{baor+99}, where this is a natural ansatz. In this work, we have formulated an envelope Hamiltonian and applied it successfully to the opposite limit, namely envelopes which vary fast as compared to the optical cycle and even the natural time scale of the bound electron. 

The success of the envelope Hamiltonian is rooted in the fact that (i) overall the ionization is small allowing the formulation of an adiabatic time-dependent perturbation theory and (ii) that different processes which produce photo electrons can be separated according to the number of photons absorbed: From non-adiabatic transitions requiring no photons (only a fast changing laser envelope) to single or eventually multiple photon ionization contributions. Thereby, each photon leads to a blue-shift of the spectral appearance of the corresponding photo electrons by $\omega$, separating the contributions well in energy for pulses not too short. Many applications of the envelope Hamiltonian for quite different physical scenarios fulfilling the hierarchy of time scales \eq{time_hierarchy} are feasible. Given the lack of accurate approximations for the electron dynamics under ultrashort pulses, this regime appears attractive for the envelope Hamiltonian particularly in connection with the adiabatic time-dependent perturbation theory as introduced here.
 
This work was supported by the COST Action XLIC (CM 1204) and the Marie Curie Initial Training Network CORINF.

\end{document}